\documentclass[letterpaper,11pt]{JHEP3}

\usepackage{epsfig}
\usepackage{graphicx}
\usepackage{cite}

\newcommand{\be}{\begin{equation}}
\newcommand{\ee}{\end{equation}}
\newcommand{\bea}{\begin{eqnarray}}
\newcommand{\eea}{\end{eqnarray}}
\renewcommand{\thefootnote}{\fnsymbol{footnote}}
\title{
Time dependent black holes and thermal equilibration
}

\author{ Dongsu Bak\\
Physics Department, University of Seoul, Seoul 130-743, Korea\\
Email: \email{dsbak@mach.uos.ac.kr
}}
\author{Michael  Gutperle\\
Department of Physics and Astronomy, UCLA, Los Angeles,
CA 90055, USA\\
Email: \email{gutperle@physics.ucla.edu}}
\author{Andreas Karch\\
 Department of Physics, University of Washington, Seattle,
    WA 98195--1560, USA\\
Email: \email{karch@phys.washington.edu}}

\abstract{
We study aspects of a recently proposed exact time dependent black
hole solution of IIB string theory using the AdS/CFT correspondence.
The dual field theory is a thermal system in which initially a
vacuum density for a non-conserved operator is turned on. We can see
that in agreement with general thermal field theory
expectation the system
equilibrates: the expectation value of the non-conserved operator
goes to zero exponentially and the entropy increases. In the field
theory the process can be described quantitatively in terms of a
thermofield state and exact agreement with the gravity answers is
found.
}

\keywords{AdS/CFT Correspondence, Black Holes}
\begin{document}
\baselineskip 16pt
\def\nn{\nonumber}

\renewcommand{\thefootnote}{\arabic{footnote}}
\setcounter{footnote}{0}

\section{Introduction}

Recently, two of the present authors together with S. Hirano proposed a family of time
dependent black hole solutions in 3 and 5 spatial dimensions that
can be embedded into type IIB string theory \cite{SHirano}. In this
paper we further interpret these solutions using the AdS/CFT
correspondence \cite{Maldacena:1997re,Witten,Gubser}, which relates the properties of the gravitational
system to those of a field theory in one lower dimension. In the
field theory we will demonstrate that the process we study
corresponds to thermal equilibration. The black hole solution
corresponds to a thermal bath in the field theory. On top of this
thermal background initially the Lagrange density, which for a
Maxwell field would be proportional to $\vec{E}^2 - \vec{B}^2$, has
a non-trivial expectation value. While the energy density $\vec{E}^2
+ \vec{B}^2$ is a conserved quantity, the Lagrange density is not
and one would expect the system to thermalize, eventually
partitioning the energy equally between electric and magnetic
fields. On general grounds this return to equilibrium should be
exponential with a characteristic thermalization time
$\tau_{therm.}$. Since the process is dissipative, the entropy
should increase during thermalization. All these expectations are
born out by explicit calculations.

Studying thermalization in any strongly coupled system is inherently
difficult. The quark gluon liquid produced at RHIC, where the
observed very short thermalization time clashes with weak coupling
expectation, is an example where this issue is of practical
interest. The AdS/CFT correspondence
is a nice tool to analytically study certain
solvable strongly coupled field theories and will hopefully serve to
build our intuition about thermalization in strongly coupled
systems. Earlier studies of thermalization using
the AdS/CFT correspondence
 were either
limited to small fluctuations around a thermal configuration, see
e.g. \cite{Kovtun:2005ev, Janik:2006gp,Friess:2006kw}, or
approximate late time solutions \cite{Bak:2006dn}. In contrast, our
gravity solution is an exact answer for all times with a large
initial perturbation. The particular field theory we study, 
$N=4$ super-Yang-Mills on a compact hyperbolic space at a fixed
temperature, is too remote from the QCD fireball at RHIC to be of
direct experimental relevance, but we find it remarkable that in
this simple system the full thermalization process can be mapped out
using the AdS/CFT correspondence.
The thermalization time we find, $\tau_{therm.} =
\frac{1}{2 \pi T}$, has appeared before in other studies
\cite{Kovtun:2005ev,Herzog:2006gh,Herzog:2006se} where it played the
role of a limiting value.
For $T= 300\ \rm MeV$, this would give $\tau_{therm.} \sim 0.1 \ \rm 
fm/c$. 
Reassuringly this is much faster than one would expect from a 
perturbative analysis.

Beyond the interest in studying thermalization our result is of
notice in that it does not exhibit any sign of a Poincar\'e
recurrence. This result demonstrates a clash between unitarity and
the strict large number of color limit in the field theory. Such a
clash is not unexpected, see \cite{MaldacenaE,Festuccia:2006sa} for
recent discussions.

In the following section we will review the time dependent black
hole solution of \cite{SHirano}. In section 3 we calculate the
properties of the dual field theory using the standard AdS/CFT
dictionary. In section 4 we use the formalism of thermofield theory
to construct a particular entangled state that encodes the exact
density matrix of the dual field theory. In section 5 we use this
thermofield state to calculate the expectation values of the
Lagrangian and the Hamiltonian to leading order in the deformation
parameter from the field theory side and establish agreement with
the gravity calculation. In section 6 we discuss the implications
our result has for the question of unitarity of the dual field
theory.

\section{Time dependent black holes}

We begin with a scalar Einstein gravity described by
\begin{equation}
I=  {1\over 16\pi G}\int d^d x \sqrt{g}\left(R-
g^{ab}\partial_a\phi
\partial_b\phi + (d-1)(d-2)
\right)\ ,
\label{daction}
\end{equation}
%
where the spacetime dimension $d$ is greater than or equals to
three. As shown in Refs.~\cite{SHirano,Hirano}, any solution of the above
can be embedded into the type IIB supergravity for $d=3$ and $d=5$.
The $d=3$ ($d=5$) solution  describes the deformation of
AdS$_3\times S^3$ (AdS$_5\times S^5$) geometry. Note that in three
dimensions, $\phi$ is the IIB dilaton whereas $\sqrt{2}\phi$
corresponds to the dilaton for $d=5$. We shall denote the  dilaton
field by $\tilde\phi$ including this extra normalization factor. The
AdS radius is denoted by $l$, which we set to be unity.

The black hole solution describing a non-equilibrium thermal
system was obtained  in Ref.~\cite{SHirano}
using the method of so called Janus construction \cite{Hirano}.
Its metric ansatz is taken as
\begin{equation}
ds^2 = f(\mu) ( d\mu^2 + ds^2_{d-1} )
\ee
where $(d-1)$ dimensional metric $\bar{g}_{pq}$
satisfies
\be
\bar{R}_{pq} = -(d-2)\bar{g}_{pq}\,.
\ee
The Einstein equations are reduced to
 \begin{equation}
f'f' = 4 f^3 -4 f^2+{4\gamma^2\over (d-1)(d-2)} f^{4-d}\ ,
\label{einsteinc}
\end{equation}
and the scalar equation is integrated once giving
\begin{equation}
\phi'(\mu) = {\gamma\over f^{d-2\over 2}(\mu)}\,,
\label{dilac}
\end{equation}
where $\gamma$ is the integration constant
responsible for the Janus deformation. For $\gamma^2 \ \le\  \gamma_c^2 $ with
\be
\gamma_c^2= (d-2)\left({d-2\over d-1}\right)^{d-2}\!\!\!\!,\ee this
can be solved by the integral \cite{SHirano}
\begin{equation}
\mu_0\pm \mu=\int^\infty_{f} {dx\over 2
\sqrt{x^3-x^2+{\gamma^2\over (d-1)(d-2)} x^{4-d}}}\,,
\end{equation}
where $\mu_0$ is chosen such that $\mu=0$ at the turning
point. 
One may show that $\mu_0 \ge \pi/2$.
Defining $\phi_\pm$ by
\be
\phi_\pm = \phi(\pm \mu_0)\,,
\ee
one finds that
\be
\phi_+ - \phi_-  = \int^{\mu_0}_{-\mu_0} d\mu{\gamma\over
f^{d-2\over 2}(\mu)}\,.
\ee
Now the trick is to take the metric $\bar{g}_{pq}$
as the cosmological form,
\begin{equation}
ds^2_{d-1}=
-d\tau^2 + \cos^2\tau ds^2_\Sigma
\end{equation}
where $ds^2_\Sigma$ is describing the compact, smooth,
finite volume Einstein space metric in $(d-2)$
dimensions satisfying
$R^\Sigma_{kl}= -(d-3) g^{\Sigma}_{kl}$.
The coordinate $\tau$ is ranged over $[-\pi/2, \pi/2]$.
For $d=3$ case, the $\Sigma$ space corresponds to
a circle $S^1$ and, for higher dimensions, the space
can be given by
 the quotient of
the hyperbolic space $H_{d-2}$ by a discrete subgroup of
the hyperbolic symmetry group, $SO(1,d-2)$.

In summary the metric for the time dependent black hole is given by
\be
ds^2 = f(\mu) ( d\mu^2 -d\tau^2 + \cos^2\tau ds^2_\Sigma)\,.
\ee
The above form of the metric is suitable for the drawing of the
 Penrose diagram.
The $\tau$ and $\mu$ coordinate can be used to represent the global
structure of the spacetime. Since $\mu_0\ge \pi/2$, the diagram is
no longer a square but a rectangle elongated in the horizontal direction.
The $\pm 45^\circ$ lines describe the future and past horizons.

\begin{figure}[ht!]
\centering \epsfysize=9cm
\includegraphics[scale=0.8]{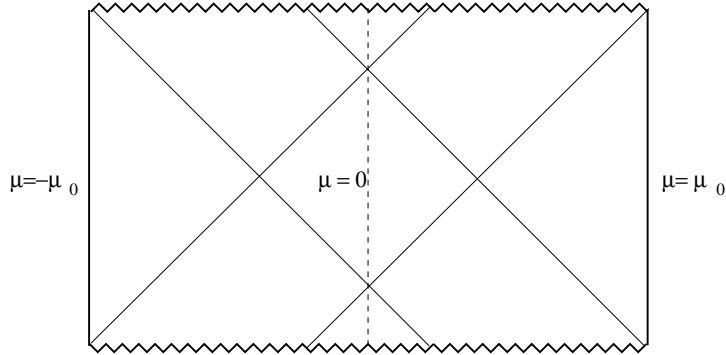}
\caption{\small Penrose diagram for the time dependent  black
hole. The $\tau$ ($ \in [-\pi/2,\pi/2]$) coordinate runs
vertically upward and $\mu$ ($ \in [-\mu_0,\mu_0]$) to the right
horizontally.}
\end{figure}

For instance, the future horizon extended from the future infinity
on the right hand side corresponds to a line $\mu-\mu_0 =\tau
-\pi/2$. Representing the volume of $\Sigma$ space by ${\cal
V}_\Sigma$, the future-horizon area is given by
\begin{eqnarray}
A(\tau)= {\cal V}_\Sigma \Big[\cos(\tau) f^{1\over 2}(\mu_0
+\tau-\pi/2)\Big]^{d-2} \,.
\end{eqnarray}
One can check that the area is monotonically increasing as a
function of $\tau$ starting from zero at $\tau = -\pi/2$ reaching
the maximal value ${\cal V}_\Sigma$ at $\tau=\pi/2$. This is
consistent with the area theorem of the black hole horizon.

To understand the geometry a little better it is instructive to look
at the undeformed case. If the deformation parameter $\gamma$ is
zero, the scalar field becomes a trivial constant and the metric
\be
ds_0^2 = {1\over \cos^2\mu }( d\mu^2 -d\tau^2 + \cos^2\tau
ds^2_\Sigma)
\ee
describes the static black hole. The Penrose diagram for this case
becomes a square and the physics of corresponding black hole is
studied in Refs.~\cite{Mann,Banados7,Brill,Vanzo,Mann7,Cr,
Birmingham,Emparan,Buchel}. Locally this
space is just Anti-de Sitter space. The only difference to global
AdS is that we had to perform an orbifold of the hyperbolic space
$H_{d-2}$ to obtain the compact manifold $\Sigma$.

By the coordinate transformation,
\be
\tanh t ={\sin\tau\over \sin\mu}, \ \ \
r = {\cos\tau\over \cos\mu}\,,
\ee
the metric of the undeformed black hole may be brought to the form
of the BTZ type \cite{Banados},
\be
ds_{eq}^2 =  -(r^2-1)dt^2 +{dr^2\over r^2-1} +
r^2 ds^2_\Sigma\,.
\ee

The temperature of the black hole is $T_{eq}= {1\over 2\pi}$ which
 is in the unit of the AdS radius.
The horizon is at $r=1$ and the corresponding black hole entropy
is
\be
S_{eq} = {{\cal V}_\Sigma \over 4 G}
\,.
\ee
The mass of the black hole is evaluated as\cite{Birmingham}
\be
M_{eq} = {{\cal V}_\Sigma \over 8\pi G } \left({d-2\over d-1}\right)
\left({d-3\over d-1}\right)^{d-3\over 2}\,.
\ee
As explained in \cite{Emparan} the value $T= {1 \over 2 \pi}$ is
also special in the dual field theory on a hyperbolic space, since
this particular thermal state can be formally obtained from the
vacuum of the Einstein universe. At other values of the temperature
the dual black hole is no longer locally AdS, but instead is given
by a $k=-1$ Schwarzschild black hole with a non-trivial mass
parameter, which for the $k=-1$ slicing formally is negative for
temperatures less than the special $T= {1 \over 2 \pi}$. To find
time dependent solutions for any temperature other than $T= {1 \over
2 \pi}$ one has to turn on an explicit time dependence in the
dilaton, $\phi(\mu,\tau)$, and the metric function $f(\mu,\tau)$. While
such a solution can at least be obtained for small deformation
parameter $\gamma$ or small temperature difference $T= {1 \over 2
\pi} + \delta T$ by perturbing around the known solutions, we will
limit our analysis in this paper to the fully solvable case of $T=
{1 \over 2 \pi}$.

Since the effect of our time dependent deformation vanishes 
at late times, these
thermodynamic properties of the undeformed case will describe the 
deformed solution at late times when the system returns to equilibrium. 
The time dependent solution describes non-equilibrium physics and we 
have to carefully determine the energy density and the entropy 
independently and cannot rely on the first law of thermodynamics to 
relate them. At least the energy density can be reliably defined even in 
the time dependent context using the holographic stress-energy tensor, 
and we will determine it in the next section where we discuss the 
gauge/gravity correspondence.

\section{Correspondence}
In this section we like to discuss the physics of
 the super Yang-Mills (SYM) theory dual to the time dependent
black hole.
Since we are dealing with the deformation of the
AdS/CFT correspondence, we shall use the standard framework
given in \cite{Witten, Gubser} for the interpretation of the
geometry of the time dependent black hole. Namely the on-shell
supergravity action with given boundary sources is providing
 the standard generating functional of connected correlators of
operators dual to the sources.
In this framework, the classical geometry and the bulk spacetime
have  a natural dual  Yang-Mills theory interpretation in the planar
large $N_c$ limit.

First by choosing the conformal factor  $h^2= \cos^2\tau/f(\mu)$,
the boundary metric for the CFT is given by\cite{SHirano}
\be
ds^2_B = -dt^2 + ds^2_\Sigma\,.
\ee
There are two separated boundaries at $\mu=\pm \mu_0$
and the boundary time $t$ ($\in (-\infty, \infty)$)is related
to $\tau$ by $\tanh t = \pm \sin\tau$ respectively
at each boundary.

For $d=5$ case, the $N=4$ SYM theory on the
above boundary spacetime is the corresponding
dual system. Since the values of the dilaton on the two
boundaries are different from each other,
the corresponding CFT's of the two boundaries now become
different as a result of the deformation. The number of colors
$N_c$ agrees with each other while the 't Hooft couplings
$\lambda_\pm$ become different by the time dependent
deformation\footnote{The 't Hooft coupling $\lambda$ is related to
the string coupling and the number of color  by
$\lambda = 4\pi g_s N_c$.}.
The situation in the $d=3$ case is not much different.
The CFT Lagrangian density is proportional to the inverse of the string
coupling  by
\be
{\cal L}_\pm \ \propto\   (g^\pm_s)^{-1} = e^{-\tilde\phi_\pm}\,.
\ee
The scalar field behaves,  in the near boundary region,  as
\be
\phi \sim \phi_\pm \mp {\gamma\over d-1}
|\mu\mp \mu_0|^{d-1} + 
\cdots
\,.
\ee
We shall be discussing the behavior of the CFT from the
view point of
the right hand side boundary at which
the upper sign in the above is relevant.
The system on the left hand side can be
treated in a similar manner.
Noting that
\be
h \sim  |\mu-\mu_0| /|\cos\tau| =
 |\mu-\mu_0| \cosh t
\ee
near boundary region of  the right hand side,
the near boundary behavior
of the scalar can be presented as
\be
\phi \sim \phi_+ - {\gamma\over (d-1)\cosh^{d-1} t}
h^{d-1} + \cdots
\,.
\ee
Since the operator dual to the dilaton is the
CFT Lagrange density, one is led to
\be
\langle {\cal L} \rangle = {\tilde\gamma\over 8 \pi G}
{1\over \cosh^{d-1}t}\,,
\label{expectation}
\ee
where $\tilde \gamma$ equals to $\sqrt{2}\gamma$ for $d=5$ and to
$\gamma$ in $d=3$ as long as we use the standard definitions of the
field theory Lagrangians. For other dimensions no string theory
embedding of the Janus geometries and no dual field theory have been
proposed so far, so we will assume that in all those cases the
Lagrange density of the dual field theory is scaled in such a way
that $\tilde \gamma = \gamma$.

By studying the near boundary behavior of the metric tensor,
one can obtain the expectation value of the boundary energy
momentum tensor. We follow the holographic renormalization methods
in Ref.~\cite{Skenderis} and the result is
\bea
\langle T_{00}\rangle &=&
{1\over 8\pi G } \left({d-2\over d-1}\right)
\left({d-3\over d-1}\right)^{d-3\over 2}\nonumber\\
\langle T_{ij}\rangle &=&
{1\over 8\pi G } \left({1\over d-1}\right)
\left({d-3\over d-1}\right)^{d-3\over 2} h_{ij}\,,
\label{enmom}
\eea
where $h_{ij}$ is the metric tensor for the $\Sigma$ space. The
total energy of the system $E$ agrees with the equilibrium value
$M_{eq}$. The result is consistent with the tracelessness condition
of the energy momentum tensor, i.e. $T^\mu\!_\mu=0$, which is due to
the conformal symmetry.

There is one important clarification about the
above computation.
When we calculate the operator expectation values or more
generally
correlators using the
on-shell gravity action, we are not using the Lorentzian
geometries. If we were working in the
Lorentzian signature, we would be troubled with the
singularities in solving the gravity equations. Instead one
 computes correlators using Euclidean geometries. As will be
shown in the next section, the Euclidean black hole geometry is
perfectly smooth and regular everywhere. There is no notion of
horizon there. Hence the gravity equation with specified
boundary sources is well defined and gives a unique
solution. Then using the corresponding on-shell action,
one can compute the Euclidean correlators of the boundary CFT
operators. One then gets the Lorentzian-signature
correlators  by the Wick rotation (in the boundary CFT) or by
an appropriate analytic
continuation (on the geometric side).

\ From the above behavior of the expectation values, it is clear
that we are dealing with a quantum system whose state is not
stationary. The system is homogeneous over the finite-volume
$\Sigma$ space, which explains the time independence of the boundary
energy momentum tensor. As the black hole possesses the $Z_2$ time
reversal symmetry $\tau \rightarrow -\tau$, the same is true for the
boundary system, which is symmetric under the time reversal $t
\rightarrow -t$, too.

The system starts  at $t=0$ from an out-of-equilibrium state where
the kinetic  energy  differs from the potential energy. Then this
out-of-equilibrium situation settles down as time goes by reaching
exponentially the equilibrium state where the kinetic energy equals
to the potential energy. The exponential approach of the equilibrium
can be seen clearly from the late time behavior of the expectation
value of the Lagrange density.
Recalling that we work at a temperature of $\frac{1}{2 \pi}$ in
units where the curvature radius of the hyperbolic space is 1, the
thermalization time can be written as $\tau_{therm.} = \frac{1}{2
\pi T}$. The final entropy of the system is given by $S= S_{eq}$,
which should be larger than the initial entropy $S_0$ at $t=0$.

A few comments are in order. Since we are dealing with a
thermal equilibration process, the first law
does not have to  hold. Namely $TdS\neq dE + p dV$.
Since the total energy and volume of the system are constant in time,
the right hand side is zero whereas the left hand side (if defined)
cannot be vanishing because of the change of the entropy.
This is not a problem since the system is not in a
quasi-equilibrium state.
Even the second law is not working since the system has the $Z_2$
symmetry and, the entropy should be decreasing as a function of
time for $t < 0$ reaching the minimal value at $t=0$.
However for the finite entropy system this kind of fine tuning
at $t=-\infty$  is not totally impossible.

\ From the geometry we have already computed the horizon area
as a function of $\tau$ which is related to the
boundary time by $\tanh t=\sin \tau$. Along the future horizon
on the right hand side, the horizon area is monotonically
increasing as we discussed before.
But for the time dependent case, we do not have a formalism to
relate the horizon area to the entropy or some other quantity
of the boundary CFT. Similarly the temperature as a function of
time cannot be computed from the geometry because
there is no notion of the periodicity of the Euclidean thermal
circle for the time dependent case. Since the system is not in
a thermal equilibrium, we do not know how to define
the temperature of the system  either.

Finally let us describe one failure of the geometric
description. The late time behavior of the expectation value of
the Lagrange density computed from geometry appears extremely
natural
from the physics view point of the thermal equilibration.
However this behavior is not consistent with the quantum Poincar\'e
recurrence theorem\cite{A11,A12,A13,A4}. The theorem states that
for any quantum system  the wave function or expectation values
of operators of the system will continuously return arbitrarily
closely to their initial values in a finite amount of time
once the spectrum is discrete.
Since the boundary  quantum system  indeed has the
discrete spectrum, one can see that there is a failure
in the geometric description once we accept the AdS/CFT
correspondence.

As discussed Ref.~\cite{MaldacenaE}, this failure of the geometrical
description
is of order $e^{- a S_{eq}}$ with some order-one   constant $a$.
The discrepancy then becomes relevant around $t \sim  a S_{eq} l$. Since
$S_{eq} \propto {N_c^2} = {\lambda^2/( 4\pi g_s)^2}$ for $d=5$,
one can see that the effect is nonperturbative in its nature.

Finally as discussed in Refs.~\cite{Witten:1999xp,Buchel}, the $d=5$
geometry shows a nonperturbative instability corresponding to
$D3-\overline{D3}$ pair creation. The same instabilities are present
in the $N=4$ SYM theory on the $\Sigma$ space which is negatively
curved. Hence the correspondence is still working in this respect.
The instabilities can be suppressed by taking the volume ${\cal
V}_\Sigma$
 large or the string coupling
$g_s$ small. Since we are in the decoupling large $N_c$ limit
with $\lambda$ fixed, the instabilities can be ignored and do not affect
 any discussions  above.

\section{Construction of the thermofield state}

The time dependent black hole solution allows
 an analytic continuation,
$\tau =-i \tau_E$, leading to
the Euclidean geometry,
\be
ds_E^2 = f(\mu) ( d\mu^2 +d\tau_E^2 + \cosh^2\tau_E ds^2_\Sigma)
\label{euclidean}
\ee
with the scalar field $\phi(\mu)$ intact. $\tau_E$ is ranged over
$(-\infty,\infty)$. The Euclidean geometry is smooth everywhere and
has a  boundary. The conformal shape of the $(\mu, \tau_E)$ space is
a disk  as the case of the usual time-independent
 black hole.
In this section we shall provide the physical interpretation of the
above Euclidean geometry in terms of thermofield dynamics
\cite{Schwinger,Bakshi,Keldysh,Takahasi}.

For the $4d$ Poincar\`e-invariant field theories,
their instanton solution  possesses $O(4)$ invariance and let us
take $t_E=0$ as the  fixed point of the $Z_2$
symmetry $t_E\rightarrow -t_E$ where $t_E$
is the Euclidean time. At this point the time derivative of
fields vanish again due to the $Z_2$ symmetry.
This $t_E=0$ field configuration may be interpreted as
an initial configuration from which the Lorentzian dynamics follows.
The subsequent Lorentzian dynamics for $t\ge 0$
can be obtained
from the instanton solution
 by analytic continuation\cite{Callan}.
At $t_E=t=0$, the Lorentzian  and the Euclidean configurations agree
with each other and the time derivatives (velocities) of both fields
vanish, which helps them join smoothly. Thus at least
semi-classically we conclude that the Euclidean solution provides an
initial state for the Lorentzian time evolution.

In case of geometry, this procedure corresponds to the
Hartle-Hawking construction of the wave function\cite{Hartle}. In
our problem we shall follow the proposal of Ref.~\cite{MaldacenaE}
to construct the corresponding thermofield initial state. Namely we
patch the half of Euclidean geometry sliced at $\tau_E=0$ to the
upper half of the Lorentzian solution sliced at $\tau=0$. Since the
geometry involves two boundaries, the corresponding Hilbert space
consists of ${\cal H}= {\cal H}_+ \times {\cal H}_-$. Unlike the
conventional thermofield formalism, the two Hamiltonians $H_+$ and
$H_-$ differ from each other, which is responsible for the time
dependence of a single boundary description as we will show below.
According to the proposal in Ref.~\cite{MaldacenaE}, the Euclidean
geometry defines a boundary Hamiltonian and allows a Euclidean
boundary time evolution, which will determine the initial
thermofield state by
\be
|\Psi\rangle = {1\over \sqrt{Z}} \sum_{m n} \langle E^+_m |U|E^-_n\rangle
 \,\,|E^+_m\rangle\times
|E^-_n\rangle
\ee
where $Z$ is the normalization factor.
The
Euclidean evolution operator $U$ is given by
\be
U = T \exp\Big[-\int^{s_+}_{s_-} d s \, H(s)\Big]
\ee
where $s$ is the boundary Euclidean time
with $s_\pm$ denoting the two boundary times
at $\tau_E=0$.
Since the two boundary Hamiltonians differ from each other,
$H(s)$ becomes  time dependent. In this respect, the above is a small
generalization of the Maldacena's proposal but this  naturally
follows  from  the fact that the Euclidean boundary Hamiltonian
is now time  dependent.

The boundary of the metric (\ref{euclidean}) can be identified by the
fact that the scale factor $f(\mu) \cosh^2\tau_E$ is
infinitely large on the boundary. The boundary is then
$\mu=\pm\mu_0$ and $\tau_E = \pm \infty$. In $(\mu, \tau_E)$
space, $(\mu ,\pm\infty)$ become two points
on the boundary, which are antipodal.
At these points, the segments $\mu=\pm\mu_0$ are joined to form
a complete circle. The conformal shape of $(\mu, \tau_E)$ space
 is depicted in Figure 2.

\begin{figure}[ht!]
\centering \epsfysize=9cm
\includegraphics[scale=1]{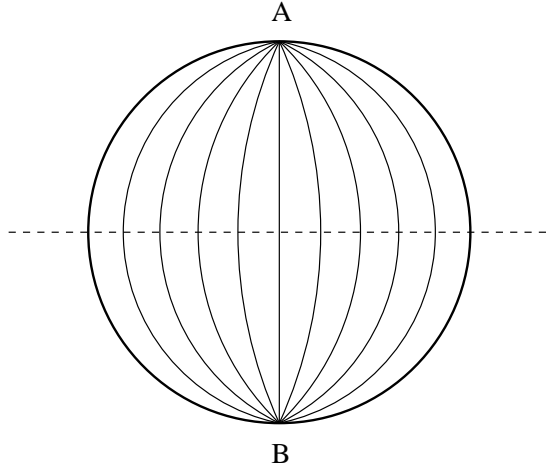}
\caption{\small The conformal diagram of the Euclidean
solution in ($\mu$, $\tau_E$)
space. The curves represent constant $\mu$ lines.
Along the curves, $\tau_E$
runs from $-\infty$ at $B$ to $+\infty$ at $A$
in the upward direction.
The right (left) half of boundary corresponds
to $\mu= \mu_0$ ($\mu=-\mu_0$). The dotted line is the
$\tau_E =0$ line and the lower half is used to construct
the thermofield state.
}
\end{figure}

In the previous section, we have introduced the boundary time
$t$ by $\tanh t = \pm \sin\tau$ respectively for
$\mu= \pm \mu_0$. By the analytic continuation $t= -i t_E$, the
Euclidean boundary time $t_E$ is related to
$\tan t_E = \pm \sinh \tau_E$. For $\mu=\mu_0$ ($\mu=-\mu_0$),
$t_E$ is chosen to run
over $[\pi/2, 3\pi/2]$ ($[-\pi/2,\pi/2]$) from $B$ ($A$) to
$A$ ($B$).  In the lower half part of the Euclidean geometry,
the boundary time ranges over $[0, \pi]$. The boundary
Hamiltonian is identified as
$H_\pm$ (that is obtained from ${\cal L}_\pm$
of the previous section)
respectively for $\mu=\pm\mu_0$. The evolution operator
$U$ then becomes
\be
U= e^{-{\pi\over 2}  H_+} e^{-{\pi\over 2} H_-}
=e^{-{\beta_{eq}\over 4} H_+} e^{-{\beta_{eq}\over 4} H_-}\,,
\ee
where we have used $\beta_{eq} \equiv 1/T_{eq} = 2\pi $.

Therefore the thermofield initial state becomes
\be
|\Psi\rangle = {1\over \sqrt{Z}} \sum_{m n}
\langle E^+_m |E^-_n\rangle e^{-{\beta_{eq}\over 4}( E^+_m + E^-_n) }
 \,\,|E^+_m\rangle\times
|E^-_n\rangle\,.
\label{state}
\ee
The state becomes the usual one if $H_+ = H_-$.

Since two boundary CFT's  are independent, the
generic time
evolution involves two boundary times $t_+$ and $t_-$
with the Hamiltonians $H_+$ and $H_-$ respectively. More
explicitly,
\be
|\Psi(t_+,t_-)\rangle  = e^{-i(t_+ H_+\times I + t_- I\times H_-)}
|\Psi(0,0)\rangle \,.
\ee
If $H_+=H_-$, the thermofield initial state is invariant
under $\tilde{H}= H_+\times I - I\times H_- $. But for
the present
case with deformation, we do not have this
symmetry any more and this will be the reason for the
time dependence of the thermal system.

The
density matrix $\rho_+$ for the boundary system on the right hand side
 is given by
\be
\rho_+ = {\rm tr}_- |\Psi\rangle \langle \Psi|
\ee
where $ {\rm tr}_\pm $ denotes the trace over
the ${\cal H}_\pm$ Hilbert space.
If $O_+$ is any operator defined in the
${\cal H}_+$ space, the thermal
expectation values are defined by
\be
\langle O_+ \rangle =
 {\rm tr}_+ \rho_+ O_+ \,.
\ee
This description of the system by the density matrix
provides us with the single boundary view.

The density matrix $\rho_+$  is no longer
commuting with $H_+$ and, hence, time dependent.
Then
the expectation value $\langle O_+ \rangle$ is
in general
time dependent, which is consistent with
our result of the previous section for $O_+ = {\cal L}$.

Note that the pure state expectation value,
$\langle \Psi | O_+ |\Psi\rangle$,  is also giving the
expectation operator $\langle O_+ \rangle$.
Therefore we conclude that the
above density matrix (or the thermofield state)
of the boundary CFT is corresponding to the geometry of the
time dependent black hole if one ignores the failure
of the previous section   that is
nonperturbative.


\section{Check for the thermofield state}

In order to verify our proposal for the thermofield state we would
like to compute the expectation value of the Lagrangian and the
Hamiltonian in the field theory and compare with the supergravity
answers in equations (\ref{expectation}) and (\ref{enmom}) to
leading order in the deformation parameter $\gamma$.
As in \cite{SHirano,Karch} one can employ the techniques of
conformal perturbation theory to calculate field theory expectation
values (suitably generalized to the
non-trivial thermofield state) in a power series in $\gamma$.
The expectation value of the Lagrangian in (\ref{expectation}) is
exactly linear
 to all orders
in $\gamma$.
The
expectation value of the
Hamiltonian that one gets by integrating the energy density $T_{00}$
in (\ref{enmom}) is the exact answer as well.
Independent of $\gamma$ one finds
the equilibrium values, so that the leading correction to $ \langle
H \rangle $ is zero in the gravity calculation.

To compute the expectation values in the field
theory we start from the
thermofield state (\ref{state}).
Note that in the following we specialize to the case $d=3$.
For the
expectation value of the Lagrange density
$\langle {\cal L}_+ (t,0)\rangle$
we get
\be
\langle {\cal L}_+ (t,0)\rangle
= {1\over Z} {\rm tr}  {\cal L}_+ (t,0)
 e^{-{\beta_{eq}\over 4} H_+} e^{-{\beta_{eq}\over 2} H_-}
e^{-{\beta_{eq}\over 4} H_+}
\ee
where  $\beta_{eq} = 2\pi$ in our case.
Let $H_- - H_+ = \delta H$ and we expand
$e^{- \pi (H_++ \delta H)}$ by
\be
e^{- \pi (H_++ \delta H)} = e^{- \pi H_+}
- \int_0^\pi d \tau  e^{-(\pi-\tau)H_+} \delta H e^{-\tau H_+}
+ \cdots
\ee
Then the leading term
\be
{\rm tr} \, {\cal L}_+  (t, 0) e^{-2 \pi H_+ }
\ee
is vanishing.
 The remaining contribution gives
\bea
\langle {\cal L}_+ (t,0)\rangle =
-\int^{\pi\over 2}_{-{\pi\over 2}} d \tau
{1\over Z}
{\rm tr} \,{\cal L}_+ (t,0)\,\,
\delta H( -i(\tau-\pi)) e^{-2\pi H_+} +\cdots
\eea
This can be arranged as\footnote{
In a Hamiltonian framework the propagator
can be given a path integral definition
$\langle x'| e^{- \tau H'}|x \rangle
= \int^{x(\tau)=x'}_{x(0)=x} {\cal D}x e^{- S'_E}$
where $S_E'$ is the Euclidean action
$ S_E' =
(1+ \delta) S_E
=-(1+ \delta) \int^{\tau}_0 ds {\cal L}(-is)$
and $S_E$ is the Euclidean action in the undeformed
theory.
Then
$$\langle x'| e^{- \tau H'}|x \rangle
= \int^{x(\tau)=x'}_{x(0)=x} {\cal D} x e^{- S_E}
(1+ \delta \int^{\tau}_0 {\cal L}(-is))+ \cdots
=\langle
x'| e^{- \tau H}|x \rangle + \delta \langle
x'|  e^{-\tau H} \int_0^{\tau} ds \hat{{\cal L}}(-is)|x \rangle
+ \cdots$$
where $\hat{{\cal L}}(-is) = e^{sH} (K-V) e^{-sH}$ with $H= K+V$.
$K$ is the kinetic energy operator while $V$ is for
the potential operator.
So the perturbation is really in terms of
the Lagrange operator ($K-V$) with imaginary time
evolution.}
\be
\langle {\cal L}_+ (t,0)\rangle =
 (e^{\phi_+}/e^{\phi_-} -1)
\int^{\pi\over 2}_{-{\pi\over 2}}d\tau
\int^{2\pi}_0 d\theta
\langle {\cal L}_+ (t, 0) {\cal L}_+ (-i(\tau-\pi)
 , \theta) \rangle_{\gamma=0}\,.
\ee
Now let us use the formula
\be
\langle {\cal L}_+ (t_1, \theta_1) {\cal L}_+ (t_2
 ,\theta_2) \rangle_{\gamma=0}
= {1\over 16\pi G}{4\over \pi} {1\over 4}\sum^\infty_{m=\infty}
{1\over [\cosh(t_1-t_2)- \cosh(\theta_1-\theta_2+2\pi m)-i\epsilon]^2}
\label{kk}
\ee
from equation (2.5) of Ref.~\cite{MaldacenaE}.
We have fixed the normalization by
comparing to the standard AdS/CFT result for the
2-point function for the operator $O$ dual to a scalar
with kinetic term $-\frac{\eta}{2} (\partial \phi)^2$
\be\langle O(x) O(0) \rangle = \eta \frac{2 \Delta - d}{\pi^{d/2}}
\frac{\Gamma(\Delta)}{\Gamma(\Delta-d/2)} \frac{1}{x^{2\Delta}}=
\frac{1}{16 \pi G} \frac{4}{\pi} \frac{1}{x^{4}}
\ee
considering
the limit $t_1\rightarrow t_2$ and
$\theta_1\rightarrow\theta_2$. The factor $1/4$ is introduced
to cancel the square of the coefficient
$1/2$ of $\cosh x -1 ={1\over 2} x^2 + \cdots$.

Therefore one finally has
\be
\langle {\cal L}_+ (t,0)\rangle =
 {1\over  16\pi G }{2\gamma \over \pi}
\int^{\pi\over 2}_{-{\pi\over 2}} d\tau
\int^{\infty}_{-\infty} d\theta
{1\over [\cosh(t+i \tau)+ \cosh\theta]^2}
\ee
where we have used $(e^{\phi_+}/e^{\phi_-} -1)=2\gamma+
\cdots $.
To evaluate the integral, we first note that
\be
\int^{\pi\over 2}_{-{\pi\over 2}} {d\tau \over a \cos\tau+ i
b\sin \tau + c}
=2 {\tan^{-1} \left( {c-a +ib\over \sqrt{c^2-a^2+b^2}}
\right) + \tan^{-1}\left({c-a -ib\over \sqrt{c^2-a^2+b^2}}\right)
\over \sqrt{c^2-a^2+b^2}}
=
2 {\tan^{-1} \left( {\sqrt{c^2-a^2+b^2}\over a}
\right)
\over \sqrt{c^2-a^2+b^2}}
\ee
where, for the last equality, we use
$\tan (A+B)= (\tan A+\tan B)/(1-\tan A\tan B)$.
Then
\bea
\int^{\infty}_{-\infty} d\theta
\int^{\pi\over 2}_{-{\pi\over 2}} d\tau
{1\over [\cosh(t+i \tau)+ \cosh\theta]^2}
&&=4 \int^{\infty}_{0} d\theta \left(-d\over \sinh\theta
d\theta \right)
{\tan^{-1} \left( {\sinh\theta \over \cosh t}
\right)
\over \sinh\theta}\nonumber\\
&&
= {-4\over \cosh^2 t} \int^\infty_0 {dw\over w} (\tan^{-1} w/w)' \,.
\eea
The definite integral is evaluated as
\be
 -2\int^\infty_0 {dw\over w} \left({\tan^{-1} w\over w}\right)'  =\left.
{w+ (w^2-1)\tan^{-1} w\over w^2}
\right|_0^{\infty}=
\pi/2\,.
\ee
Hence we get
\be
\langle {\cal L}_+ (t,0)\rangle =
 {1\over  8 \pi G }{\gamma \over \cosh^2 t}\,,
\label{exp1}
\ee
in complete agreement with the gravity result in
(\ref{expectation}).

In an analogous fashion we can calculate the order $\gamma$
correction to the expectation value of the
the boundary energy momentum tensor
$\langle T_{\mu\nu}
\rangle$.
The
calculation proceeds as above, however this time the relevant 2-point
function is $\langle T_{\mu\nu} \cal{L} \rangle$ which vanishes (and hence
the correction vanishes, too). This is in agreement with the
supergravity result (\ref{enmom}) which states that to all orders in
$\gamma$ the expectation value of $T_{\mu\nu}$ is given by the
equilibrium answer, so in particular the order $\gamma$ correction
vanishes.

\section{Discussion}

In this paper we have investigated a time dependent black hole
solution utilizing the AdS/CFT correspondence. We have found an
exact solution of the supergravity in the large $N_c$ and large AdS
curvature radius limit, which corresponds to a large initial
perturbation of the black hole geometry and have observed an
exponential return to equilibrium.

We constructed the  thermofield state from the Hartle-Hawking
wavefunction and showed  the agreement on both sides of the duality
for the time dependent expectation value of the Lagrangian density
to lowest order in conformal perturbation theory. Hence the time
dependent black holes spacetimes, we have discussed,  yield an
interesting laboratory to study equilibration of  strongly coupled
gauge theories. A  more detailed study of these spacetimes,
including the five dimensional case, the calculation of higher point
correlation functions and higher orders in conformal perturbation
theory would be interesting. Unfortunately, the nature of the Janus
ansatz implies that the gauge theories are defined on compact spaces
of negative curvature,  which is not the case one is most interested
in for ``real world'' applications.

In Ref.~\cite{MaldacenaE} it was found that a small perturbation
around the eternal black hole leads to exponentially decaying
correlation functions, which are inconsistent with the quantum
Poincar\'e recurrence theorem and hence with unitarity. The operator
expectation values we computed from the on-shell supergravity action
are once more inconsistent with the quantum Poincar\'e recurrence
theorem. As in Ref.~\cite{MaldacenaE} this failure of the
correspondence in the large time limit can be shown to occur as a
nonperturbative effect. Therefore this failure is not a
contradiction at all. In the strict planar limit where the
evaluation in the dual supergravity theory in terms of the classical
on-shell action is valid the unitarity of the field theory is not
manifest. It only gets reinstated by considering exponentially
suppressed corrections.

\section*{Acknowledgments}
 We are grateful to Juan Maldacena for useful
discussions and conversations.  The work of DB is supported in part
by KAST, KOSEF SRC CQUeST R11-2005-021 and ABRL R14-2003-012-01002-0, 
the work of MG was supported in part by NSF grant PHY-04-56200 and work 
of AK was supported in part by the U.S. Department of Energy under Grant
No.~DE-FG02-96ER40956.


\begin{thebibliography}{99}



\bibitem{SHirano}
  D.~Bak, M.~Gutperle and S.~Hirano,
  ``Three dimensional Janus and time-dependent black holes,''
  JHEP {\bf 0702}, 068 (2007)
  [arXiv:hep-th/0701108].

\bibitem{Maldacena:1997re}
  J.~M.~Maldacena,
  ``The large N limit of superconformal field theories and supergravity,''
  Adv.\ Theor.\ Math.\ Phys.\  {\bf 2} (1998) 231
  [Int.\ J.\ Theor.\ Phys.\  {\bf 38} (1999) 1113]
  [arXiv:hep-th/9711200].


\bibitem{Witten}
  E.~Witten,
  ``Anti-de Sitter space and holography,''
  Adv.\ Theor.\ Math.\ Phys.\  {\bf 2}, 253 (1998)
  [arXiv:hep-th/9802150].

\bibitem{Gubser}
  S.~S.~Gubser, I.~R.~Klebanov and A.~M.~Polyakov,
  ``Gauge theory correlators from non-critical string theory,''
  Phys.\ Lett.\  B {\bf 428}, 105 (1998)
  [arXiv:hep-th/9802109].






\bibitem{Kovtun:2005ev}
  P.~K.~Kovtun and A.~O.~Starinets,
  ``Quasinormal modes and holography,''
  Phys.\ Rev.\  D {\bf 72}, 086009 (2005)
  [arXiv:hep-th/0506184].

\bibitem{Janik:2006gp}
  R.~A.~Janik and R.~Peschanski,
  ``Gauge / gravity duality and thermalization of a boost-invariant perfect
  fluid,''
  Phys.\ Rev.\  D {\bf 74}, 046007 (2006)
  [arXiv:hep-th/0606149].

\bibitem{Friess:2006kw}
  J.~J.~Friess, S.~S.~Gubser, G.~Michalogiorgakis and S.~S.~Pufu,
  ``Expanding plasmas and quasinormal modes of anti-de Sitter black holes,''
  JHEP {\bf 0704}, 080 (2007)
  [arXiv:hep-th/0611005].

\bibitem{Bak:2006dn}
  D.~Bak and R.~A.~Janik,
  ``From static to evolving geometries: R-charged hydrodynamics from
  supergravity,''
  Phys.\ Lett.\  B {\bf 645}, 303 (2007)
  [arXiv:hep-th/0611304].

\bibitem{Herzog:2006gh}
  C.~P.~Herzog, A.~Karch, P.~Kovtun, C.~Kozcaz and L.~G.~Yaffe,
  ``Energy loss of a heavy quark moving through N = 4 supersymmetric
  Yang-Mills plasma,''
  JHEP {\bf 0607}, 013 (2006)
  [arXiv:hep-th/0605158].

\bibitem{Herzog:2006se}
  C.~P.~Herzog,
  ``Energy loss of heavy quarks from asymptotically AdS geometries,''
  JHEP {\bf 0609}, 032 (2006)
  [arXiv:hep-th/0605191].

\bibitem{MaldacenaE}
  J.~M.~Maldacena,
  ``Eternal black holes in Anti-de-Sitter,''
  JHEP {\bf 0304}, 021 (2003)
  [arXiv:hep-th/0106112].

\bibitem{Festuccia:2006sa}
  G.~Festuccia and H.~Liu,
  ``The arrow of time, black holes, and quantum mixing of large N Yang-Mills
  theories,''
  arXiv:hep-th/0611098.


\bibitem{Hirano}
D.~Bak, M.~Gutperle and S.~Hirano,
  ``A dilatonic deformation of AdS(5) and its field theory dual,''
JHEP {\bf 0305}, 072 (2003)
[arXiv:hep-th/0304129].


\bibitem{Mann}
  R.~B.~Mann,
  ``Pair production of topological anti-de Sitter black holes,''
  Class.\ Quant.\ Grav.\  {\bf 14} (1997) L109
  [arXiv:gr-qc/9607071].

\bibitem{Banados7}
  M.~Banados,
  ``Constant curvature black holes,''
  Phys.\ Rev.\  D {\bf 57}, 1068 (1998)
  [arXiv:gr-qc/9703040].



\bibitem{Brill}
  D.~R.~Brill, J.~Louko and P.~Peldan,
  ``Thermodynamics of (3+1)-dimensional black holes with toroidal or higher
  genus horizons,''
  Phys.\ Rev.\  D {\bf 56} (1997) 3600
  [arXiv:gr-qc/9705012].

\bibitem{Vanzo}
  L.~Vanzo,
  ``Black holes with unusual topology,''
  Phys.\ Rev.\  D {\bf 56} (1997) 6475
  [arXiv:gr-qc/9705004].

\bibitem{Mann7}
  R.~B.~Mann,
  ``Topological black holes: Outside looking in,''
  arXiv:gr-qc/9709039.

\bibitem{Cr}
  J.~D.~E.~Creighton and R.~B.~Mann,
  ``Entropy of constant curvature black holes in general relativity,''
  Phys.\ Rev.\  D {\bf 58}, 024013 (1998)
  [arXiv:gr-qc/9710042].


\bibitem{Birmingham}
  D.~Birmingham,
  ``Topological black holes in anti-de Sitter space,''
  Class.\ Quant.\ Grav.\  {\bf 16}, 1197 (1999)
  [arXiv:hep-th/9808032].




\bibitem{Emparan}
  R.~Emparan,
  ``AdS/CFT duals of topological black holes and the entropy of
  zero-energy states,''
  JHEP {\bf 9906}, 036 (1999)
  [arXiv:hep-th/9906040].


\bibitem{Buchel}
  A.~Buchel,
  ``Gauge theories on hyperbolic spaces and dual wormhole instabilities,''
  Phys.\ Rev.\ D {\bf 70}, 066004 (2004)
  [arXiv:hep-th/0402174].


\bibitem{Banados}
  M.~Banados, C.~Teitelboim and J.~Zanelli,
  ``The Black hole in three-dimensional space-time,''
  Phys.\ Rev.\ Lett.\  {\bf 69}, 1849 (1992)
  [arXiv:hep-th/9204099].



\bibitem{Skenderis}
  K.~Skenderis,
  ``Lecture notes on holographic renormalization,''
  Class.\ Quant.\ Grav.\  {\bf 19}, 5849 (2002)
  [arXiv:hep-th/0209067].

\bibitem{A11}
P. Boccieri and A. Loinger, ``Quantum recurrence theorem'',
Phys. Rev. {\bf 107}, 337 (1957).

\bibitem{A12}
L.S. Schulman, ``Note on the quantum recurrence theorem'', Phys. Rev. A {\bf 18}, 2379 (1978)

\bibitem{A13}
I.C. Percival, ``Almost Periodicity and the Quantal H Theorem'', J. Math. Phys. {\bf 2}, 235 (1961).



\bibitem{A4}
  L.~Dyson, M.~Kleban and L.~Susskind,
  ``Disturbing implications of a cosmological constant,''
  JHEP {\bf 0210}, 011 (2002)
  [arXiv:hep-th/0208013].



\bibitem{Witten:1999xp}
  E.~Witten and S.~T.~Yau,
  Adv.\ Theor.\ Math.\ Phys.\  {\bf 3}, 1635 (1999)
  [arXiv:hep-th/9910245].


\bibitem{Schwinger}
  J.~S.~Schwinger,
  ``Brownian motion of a quantum oscillator,''
  J.\ Math.\ Phys.\  {\bf 2} (1961) 407.



\bibitem{Bakshi}
P. ~M.~ Bakshi and K.~ T.~ Mahanthappa, ÒExpectation Value Formalism in Quantum Field
Theory, 1,Ó  J.\  Math.\  Phys. \ {\bf 4} (1963) 4.

\bibitem{Keldysh}
  L.~V.~Keldysh,
  ``Diagram technique for nonequilibrium processes,''
  Zh.\ Eksp.\ Teor.\ Fiz.\  {\bf 47} (1964) 1515
  [Sov.\ Phys.\ JETP {\bf 20} (1965) 1018].

\bibitem{Takahasi}
  Y.~Takahasi and H.~Umezawa,
  ``Thermo Field Dynamics,''
  Collect.\ Phenom.\  {\bf 2} (1975) 55.

\bibitem{Callan}
  C.~G.~.~Callan and S.~R.~Coleman,
  ``The Fate Of The False Vacuum. 2. First Quantum Corrections,''
  Phys.\ Rev.\  D {\bf 16}, 1762 (1977).


\bibitem{Hartle}
  J.~B.~Hartle and S.~W.~Hawking,
  ``Wave Function Of The Universe,''
  Phys.\ Rev.\  D {\bf 28}, 2960 (1983).






\bibitem{Karch}
  A.~B.~Clark, D.~Z.~Freedman, A.~Karch and M.~Schnabl,
  ``The dual of Janus $((<:)$ $<-->$ $(:>))$ an interface CFT,''
  Phys.\ Rev.\ D {\bf 71}, 066003 (2005)
  [arXiv:hep-th/0407073].



\end{thebibliography}
\end{document}